\documentclass[journal=jacsat,manuscript=article]{achemso}

\usepackage[version=3]{mhchem} % Formula subscripts using \ce{}
\usepackage[usenames,dvipsnames]{pstricks}
\usepackage{epsfig}
\graphicspath{IMG/}
\usepackage{graphicx}
\usepackage{bm}
\usepackage{dcolumn}% Align table columns on decimal point
\usepackage{color,soul}

\newcommand{\A}[1]{\Hat{#1}}

\newcommand{\vp}[1]{\mathbf{#1}}

\newcommand{\abs}[1]{\left| #1\right|}

\newcommand{\gsim}{\raisebox{-0.13cm}{~\shortstack{$>$ \\[-0.07cm]
      $\sim$}}~}

\author{Behrooz Semnani}
\affiliation{Institute for Quantum Computing (IQC), University of Waterloo, Waterloo, N2L3G1, Canada}
\alsoaffiliation{Department of Electrical \& Computer Engineering, University of Waterloo, Waterloo, N2L3G1, Canada}
\email{bsemnani@uwaterloo.ca}
\author{Jeremy Flannery}
\affiliation{Institute for Quantum Computing (IQC), University of Waterloo, Waterloo, N2L3G1, Canada}
\alsoaffiliation{Department of Physics and Astronomy, University of Waterloo, Waterloo, N2L3G1, Canada}
\author{Rubayet Al Maruf}
\affiliation{Institute for Quantum Computing (IQC), University of Waterloo, Waterloo, N2L3G1, Canada}
\alsoaffiliation{Department of Electrical \& Computer Engineering, University of Waterloo, Waterloo, N2L3G1, Canada}
\author{Michal Bajcsy}
\affiliation{Institute for Quantum Computing (IQC), University of Waterloo, Waterloo, N2L3G1, Canada}
\alsoaffiliation{Department of Electrical \& Computer Engineering, University of Waterloo, Waterloo, N2L3G1, Canada}
\alsoaffiliation{Department of Physics and Astronomy, University of Waterloo, Waterloo, N2L3G1, Canada}
\title[paper]
  {Spin-Preserving Chiral Photonic Crystal Mirror}

\abbreviations{IR,NMR,UV}

\begin{document}
%%%%%%%%%%%%%%%%%%%%%%%%%%%%%%%%%%%%%%%%%%%%%%%%%%%%%%%%%%%%%%%%%%%%%
%% The manuscript does not need to include \maketitle, which is
%% executed automatically.  The document should begin with an
%% abstract, if appropriate.  If one is given and should not be, the
%% contents will be gobbled.
%%%%%%%%%%%%%%%%%%%%%%%%%%%%%%%%%%%%%%%%%%%%%%%%%%%%%%%%%%%%%%%%%%%%%
\begin{abstract}
Chirality refers to a geometric phenomenon in which objects are not superimposable on their mirror image\cite{book_Schferling2017}.  Structures made of nano-scale chiral elements can display {chiroptical} effects, such as dichroism for left- and right- handed circularly polarized light, which makes them of high interest for applications ranging  from  quantum information processing and quantum optics\cite{Lodahl2017,2018_Sciecne_Hafezi} to circular dichroism spectroscopy and molecular recognition \cite{2006_nat_Greenfield}. At the same time, strong {chiroptical} effects have been challenging to achieve even in synthetic optical media and {chiroptical} effects for light with normal incidence has been speculated to be prohibited in {lossless}, thin, quasi-two-dimensional structures \cite{2015_Springer_Oh,2007-PRA-Bai,2009_Sci_Gansel,2017_NatLS_Zhu}. Here, we report on our experimental realization of a giant {chiroptical} effect in a thin monolithic photonic crystal mirror. Unlike conventional mirrors, our structure selectively reflects only one spin state of light, while preserving its handedness, with a near unity level of circular dichroism. The operational principle of the photonic-crystal mirror relies on Guided Mode Resonance (GMR) with simultaneous excitation of leaky TE and TM Bloch modes in the photonic crystal slab. Such modes are not reliant on the suppression of their radiative losses through the long-range destructive interference and even small areas of the photonic-crystal exhibit robust circular dichroism. Despite its simplicity, the mirror strongly surpasses the performance of earlier reported structures and, contrary to a prevailed notion, demonstrates that near unity reflectivity contrast for the opposite helicities is achievable in a quasi-two-dimensional structure.  %Beside its obvious potential functionalities in chiral photonics, we expect this device to find exotic applications in cavity QED and to enable novel types of lasers.  

%This paper reports on our unprecedented experimental observation of a giant \colorbox{green}{chiroptical} effect in a thin monolithic photonic crystal mirror. The
%mirror selectively reflects one spin state of light while, unconventionally,
%preserving its handedness. The operational principle of the PC mirror relies on Guided Mode Resonance (GMR) via a simultaneous excitation of TE and TM Bloch modes  which in turn yields magneto-dipole interactions. Since such modes are not reliant on the suppression of their radiative losses through the long-range destructive interference, a finite version of the PC  would exhibit a reasonable performance. Despite its simplicity, the mirror surpasses the performance of earlier reported structures and thus in contrary to a prevailed notion, this work shows that near unity circular dichroism is achievable in a simple and monolithic slab structure.  We expect this device finds exotic applications in cavity QED and novel types of laser, beside its obvious potential functionalities in chiral photonics.  

\end{abstract}

\section{Introduction}
%Chirality refers to a geometry symmetry property which describes objects that are not superimposable on their mirror image and thereby they possess a common sense of twist \cite{book_Schferling2017}. A geometrically chiral object and its mirror image are called two enantiomorphs.  

%Apart from the geometrical considerations, chirality can manifest itself in physical and chemical properties of materials and thus it is studied in very diverse scientific disciplines (e.g. see Refs.~\cite{2017_science_Hentschel,Hendry2010} and the references therein).  In particular, the structures made of chiral elements might respond differently to the opposite spin states of light- i.e. right-handed and left-handed circularly polarized light abbreviated by RHCP and LHCP respectively-.

%Let's maybe skip this as well:
%  Interaction of light and chiral structures has been a long-explored concept in classical and quantum optics. Although somewhat ubiquitous, with examples including the wings of manuka beetles and butterflies, naturally occurring chiral compounds exhibit relatively weak \colorbox{green}{chiroptical} effects, which hinders their adoption for high-performance applications. Instead, synthetic optical media such as chiral metamaterials or their predecessors, photonic crystals (PCs), composed of subwavelength-scale building blocks carrying geometrical chirality can go far beyond the reach of any natural material, for example, by mediating cross-coupling between the electric field and magnetic field going through the medium \cite{2017_NatLS_Zhu}. 

An ultra-thin spin-preserving chiral mirror that completely reflects only one spin state of light upon normal illumination without reversing the light's handedness is a {chiroptical} structure of particular interest
%. Such a mirror cannot be implemented with conventional,
as metallic, dielectric-stack, or even Faraday mirrors flip the helicity (i.e., the spin) of light upon reflection. Additionally, Fabry-P\'{e}rot cavities made of spin-preserving mirrors would exhibit all sorts of unique properties \cite{2015_APL_Plum}, such as the distribution of the resonant field being null-free and thus the intensity of light along the cavity length remaining constant. Such self-polarizing `\textit{chiral cavities}' formed from these thin mirrors will open up tantalizing possibilities in quantum optics\cite{2015_PRL_Yoo} and opto-mechanics, with opportunities ranging from realization of novel types of gas lasers based on mesoscale fiber-integrated cavities \cite{2017_ACS_Flannery} to fundamental studies of light-matter interactions in systems such as membrane-in-the-middle coupled cavities \cite{2016_NatLS_Chen}. A particular advantage offered by self-polarizing chiral cavities is the relatively high strength of optical transitions between atomic levels coupled by circularly polarized light compared to levels coupled by linearly-polarized light. Additionally, such optical transitions with circularly polarized light are often effectively closed and allow the isolation of a two-level system from the generally complicated level structure of commonly used atoms.
 %in such self-polarizing `\textit{chiral cavities}'
 %The applications described above require an ultra-thin chiral mirror that completely reflects only one spin state of light upon normal illuminations without reversing its handedness. 

However, the realization of an ultra-thin chiral mirror that completely reflects only one spin state of light upon normal illuminations without reversing its handedness involves two outstanding challenges in nanophotonics.  First, the \textit{intrinsic chirality}, i.e. {chiroptical} effects for light of normal incidence, in quasi-two-dimensional {lossless} structures has been speculated to be prohibited \cite{2015_Springer_Oh,2007-PRA-Bai,2009_Sci_Gansel,2017_NatLS_Zhu}. As a matter of fact, the main requirement to achieve intrinsic {chiroptical} effects is a simultaneous excitation of both in-plane magnetic and electric dipole moments upon normal illuminations \cite{book_Schferling2017,2009_Sci_Gansel}, where `in-plane' refers to the quasi-2D structure intended to discriminate opposite spins of light. For this to occur, according to a long-held notion, the structure has to be composed of complicated three-dimensional chiral elements, such as helices or, alternatively, made of multi-layer patterns carrying structural chirality \cite{book_Schferling2017,2009_Sci_Gansel}. To date, several demonstrations of plasmon-assisted intrinsic {chiroptical} responses in metastructures composed of subwavelength array of 3D chiral shapes \cite{2009_Sci_Gansel,2019-ACS-Rajaei,2019-NanoLett-Yang,2017_science_Hentschel} or multilayer patterns of mirror-symmetry-broken structures\cite{2012-NatCom-Zhao,2012_PRB_Zhou,2014_NaoLett_Cui,2018_ACS_Ma,2018_NatLS_CHEN_SPIN,2018_ACS_Wu} have been reported. Although such structures may exhibit strong and wide-band chiaroptical response, their fabrication is not compatible with 2D patterning techniques \cite{2017_NanoLett_Chen}. Top-down fabrication techniques including direct laser writing have been widely employed to make 3D chiral structures with arbitrary geometries\cite{2009_Sci_Gansel}. However, such techniques are usually limited to micrometer resolution and thereby not suitable for nano-scale structures operating at visible and near infrared \cite{2019-ACS-Rajaei}. Alternative techniques such as electron beam induced deposition (EBID)\cite{2016_IOPNANO_Haverkamp} and colloidal nanohole lithography \cite{2013_AC_Nano_Frank}  are too complicated  for manufacturing large-area plasmonic structures.   

For thin, quasi-2D structures, co-excitation of the in-plane dipole moments seems too challenging to be fulfilled. In contrast to in-plane electric dipoles which can be readily excited, occurrence of in-plane magnetic dipole moments involves circulation of polarization current at the vertical cross section of the structure. This is not straightforward to accomplish in a thin structure \cite{2017_NatLS_Zhu}. On the other hand, the need for an in-plane magnetic moment is eliminated for obliquely incident light, which is commonly termed as extrinsic chirality \cite{2009_IOP_Optics_Plum,2018_ACS_Yoo}. Recently though, \citet{2017_NatLS_Zhu} reported an experimental observation of a giant intrinsic {chiroptical} effect in an optically thin metasurface composed of a periodic array of chiral gammadion-shaped meta-atoms arranged on a dielectric slab. This metasurface selectively transmits only one spin state of light while diffracting the opposite spin. The underlying physical mechanism is a selective excitation of higher-order multipoles, such as the toroidal quadrupole and magnetic octupole \cite{2016_Nat_zhelodev}. Since the primary radiation direction for the higher-order modes upon their excitation is off-normal\cite{2016_Nat_zhelodev}, the transmission at forward direction is eliminated and the structure effectively filters out the selected helicity. However, while multipole engineering can render a structure intrinsically chiral \cite{2014_NatCom_Wu,2005_PRL_KuwataGonokami}, it does not suit for designing reflective chiral structures. Moreover, as will be elucidated further, polarization conversion is a main requirement of the operation of spin-preserving mirrors and higher order multipoles are not able to introduce the desired polarization conversion at the zeroth diffraction term.

The second challenge is the requirement for handedness preservation upon normal reflection. For an ordinary mirror or any uniform dielectric interface, reversal of helicity occurs when light reflects off their surface. Several groups have reported on demonstrating of `magical mirrors' that selectively reflect  circularly polarized  light  to  its  co-circular  polarization state \cite{2015_APL_Plum,wang_2016_Metamirrors,2017_NanoLett_Kang,2016_AdvMat_Xiao,2017_APL_Jing}. Although diverse structures have been proposed, there is a commonality between the notions employed: the proposed mirrors are composed of 2D-chiral arrays arranged on top of a back metallic mirror. Judicious design of such structures would enable complete reflection of one spin state without handedness reversal, whereas the opposite spin should be completely absorbed \cite{2015_APL_Plum}. The operational principle of the magical mirrors is reliant on selective absorption of the light impinging on the metallic chiral pattern. However, absorption is fundamentally limited to $50\%$ in 2D arrays\cite{2015_APL_Plum}. It turns out that the presence of back metallic mirrors is indispensable, allowing complete absorption of the selected polarization within a round trip of propagation \cite{2015_APL_Plum}.  Despite the fact that the demonstrated mirrors can potentially lead to near unity circular dichroism \cite{wang_2016_Metamirrors}, realization of spin-preserving mirrors in monolithic planar all-dielectric structures remains a challenge.   

Here, we describe the the first experimental observation of maximum intrinsic chirality \cite{2016_PRX_Fernandez} in a truly monolithic {and lossless}  photonic crystal (PC) membrane with a chiral array of perforating holes. Our structure is designed and fabricated for operation at near infrared range. Upon normal illumination, the PC slab reflects the chosen helicity to the same state of polarization with near unity reflection coefficient while the opposite spin gets completely transmitted.  The structure overcomes the challenges outlined above by a proper hybridization of leaky Bloch modes, leading to near unity circular dichroism.  The underlying physical mechanism is Guided Mode Resonance (GMR) \cite{2002_PRD_Fan} via loosely confined TE and TM modes. The hybridization of the extremely-leaky (having low radiation quality factor) TE and TM modes entails resonant reflection of selected circular polarization to the co-circularly polarized state and the handedness preservation is attained via a spatial symmetry of the perforating holes. As a result, not only can the associated low-Q modes effectively interact with free space illumination, the need for an infinitely extended structure is largely eliminated. %The design presented in this article serves as a proof-of-concept for further developments of resonant chiral mirrors following a similar recipe. 

\section{Results}

\subsection{Design, concepts and device fabrication}

\begin{figure}[t]
\centering
\includegraphics[width=1\textwidth]{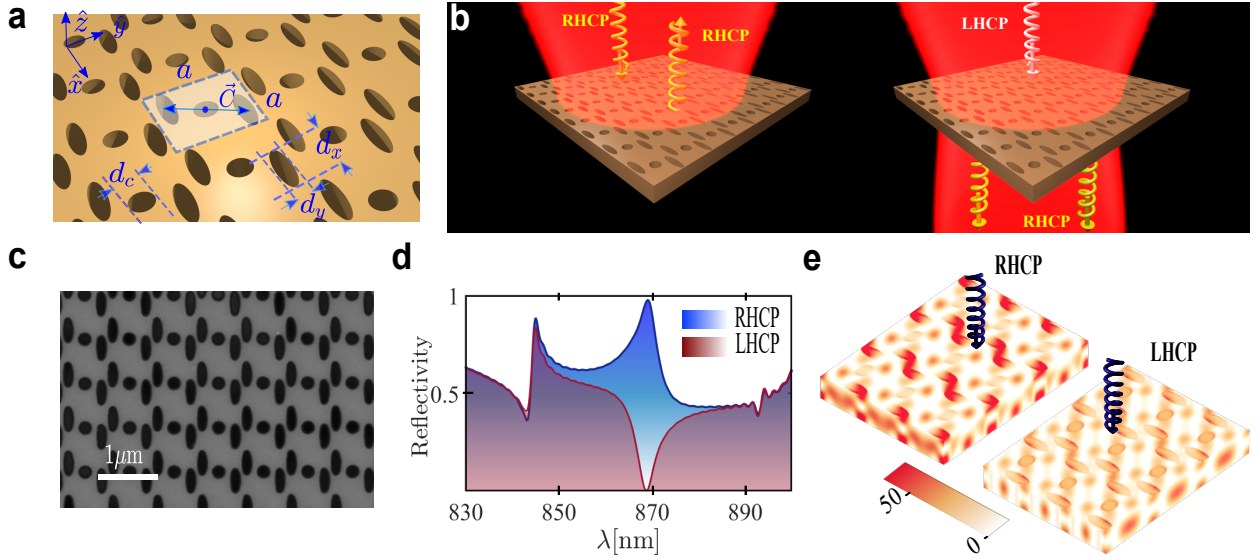}
\caption{\textbf{Schematics and simulation results} \textbf{(a)} Schematic of the chiral PC membrane and geometry definitions. The Bravais lattice is square with the lattice constant of $a=740 \mathrm{nm}$. The unit cell (shaded area) comprises a tripartite configuration of perforating holes: a circular hole at the center with a diameter of $d_c=200\mathrm{nm}$ and two elliptical holes ($d_x=420\mathrm{nm}$ and $d_y=140\mathrm{nm}$) displaced by  $\pm \vec{C}= \pm (150\A{x}+275\A{y}) [\mathrm{nm}]$ with respect to the center. The membrane is made from silicon nitride with refractive index of $n\sim 2.26$ and thickness of $\sim 309\mathrm{nm}$  \textbf{(b)} Illustration of the optical response at the designed wavelength $870 \mathrm{nm}$. The structure reflects RHCP while preserving its handedness. The opposite spin is transmitted and its handedness is reversed. \textbf{(c)} SEM  image of the fabricated device.  \textbf{(d,e)} FDTD simulation results: \textbf{(d)} power reflection spectrum  for the two spin states of the incident light and \textbf{(e)} the corresponding intensity distributions over a few unit cells.
Color axis displays normalized electric field intensity profile $\abs{E/E_0}^2$ where $E$ and $E_0$ are the induced electric field magnitude upon circularly-polarized illuminations and the magnitude of the incident field respectively. } 
\label{fig:schematics}
\end{figure}

A schematic of the designed photonic crystal mirror is displayed in Fig.~\ref{fig:schematics}\textbf{a}. The photonic crystal membrane is composed of a patterned layer of silicon nitride with a thickness of $t\sim 309\mathrm{nm}$ that optimizes operation at the target wavelength of $\lambda\sim 870\mathrm{nm}$. The design can be adjusted for other wavelengths or dielectric materials. The Bravais lattice is square-shaped with a sub-wavelength lattice constant which assures that upon free-space illumination only zeroth order diffraction will contribute to reflection and transmission in the far field. The photonic crystal membrane can thus be  regarded  as  an  effectively  homogeneous boundary. The unit cell consists of a tripartite array of perforating holes carrying chiral symmetry in the xy-plane (see Fig.~\ref{fig:schematics}\textbf{a}). Here, chirality is achieved by engineering the detailed geometry of the unit cell and the wavelength is adjusted by properly selecting the thickness of the slab.

%%%%%%%%%%%%%%%%%%%%%%%%%%%%%%%%%%%%%%
%%%%%%%%%%%%%%%%%%%%%%%%%%%%%%%%%%

 Maximum electromagnetic chirality requires preserving of the light's handedness at normal incidence which is imposed by time reversal symmetry\cite{2016_PRX_Fernandez}. In essence, the symmetry of the pattern of the unit cell dictates the basic relationship between the elements of the reflection tensor. %Sub-wavelength periodicity of the structure allows
 Employing Jones calculus within the circular basis, the reflection and transmission properties of the slab are then described by the elements of the $2\times 2$ matrices ${\vp{R}}=[\mathcal{R}_{ij}]$ and $\vp{T}=[\mathcal{T}_{ij}]$ respectively. Henceforth, the corresponding matrix elements are subscripted by $+$ and $-$  designating right-handed and left-handed circularly polarised modes, respectively. The desired reflection properties at the designed wavelength necessitates setting $\mathcal{R}_{++}=1$ (or for the appositive enantiomer $\mathcal{R}_{--}=1$ ), whereas the other three elements should be vanishingly small. This assures that the PC mirror reflects only one spin state of light without reversing its handedness, whereas the opposite spin is completely transmitted.  
It is also worth pointing out that due to its 2D nature, the structure should exhibit opposite chirality on both its sides. In stark contrast with 3D chiral objects such as a helix, where the sense of twist associated with the object is independent of the observation direction, the perceived sense of twist of a planar chiral object is reversed upon reversal of the observation direction. This, in conjunction with the time reversal symmetry, results in flipping the spin of the transmitted light (see Supplemental Material\footnote{Supplemental Material can be obtained from the authors upon request.}). Therefore, in an ideal scenario, the only nonzero element of the corresponding transmission matrix at the target wavelength is $\abs{\mathcal{T}_{+-}}=1$ (for the appositive enantiomer $\abs{\mathcal{T}_{-+}}=1$). Fig.~\ref{fig:schematics}\textbf{b} schematically illustrates the expected optical response of the PC slab at the designed wavelength.

The required structure of the reflection and transmission tensors defined above leaves us with several consequences and poses further limitations on the spatial symmetries of the perforating holes. As detailed in the Supplemental Material, only one-fold $C_1$ and two-fold $C_2$  symmetry groups accompanied with broken mirror symmetry in the xy-plane will fundamentally allow preservation of the helicity (spin) upon reflection. In other words, the necessary condition to realize such spin-preserving mirrors is to simultaneously break the $n$-fold rotational symmetry (for $n>2$) and any in-plane mirror symmetries \cite{wang_2016_Metamirrors,2007-PRA-Bai}. To also reduce the  sensitivity of the structure with respect to the angle of incidence, we elected to design a unit cell carrying two-fold rotational symmetry (see Fig.~\ref{fig:schematics}\textbf{a}).  The dimensions are initially selected using band diagram analysis and finely adjusted for maximum extinction ratio and near perfect reflection through a brute force optimization technique. The optimization is constrained to the fabrication limitations including the bridge sizes between adjacent holes as well as the smallest curvatures to be etched.  An SEM image of the fabricated photonic crystal membrane is shown in Fig.~\ref{fig:schematics}\textbf{c}.

The structure was simulated through finite difference time domain (FDTD) method using a commercial solver (Lumerical Inc.). The power reflectivity of the slab for the normal incidence of the RHCP and LHCP lights as well as the corresponding field distributions are shown in Figs.~\ref{fig:schematics} \textbf{d} and \textbf{e} respectively. The simulation results promise more than $97\%$ reflection for one spin state of light at the target wavelength, whereas the opposite spin state is almost completely transmitted. The extinction ratio can reach up to 1000 which is unprecedented among the relevant works \cite{wang_2016_Metamirrors}. This giant intrinsic chirality originates from the guided mode resonance (GMR) mediated by two extremely leaky Bloch modes across the band-edges of the PC slab. Due to bi-modal interference,
the intensity profiles at the cross-section of the photonic crystal slab (shown in Fig.~\ref{fig:schematics} \textbf{e}) are asymmetric and  symmetric for reflective and transmissive helicities respectively.

\begin{figure}[t]
\centering
\includegraphics[width=1\textwidth]{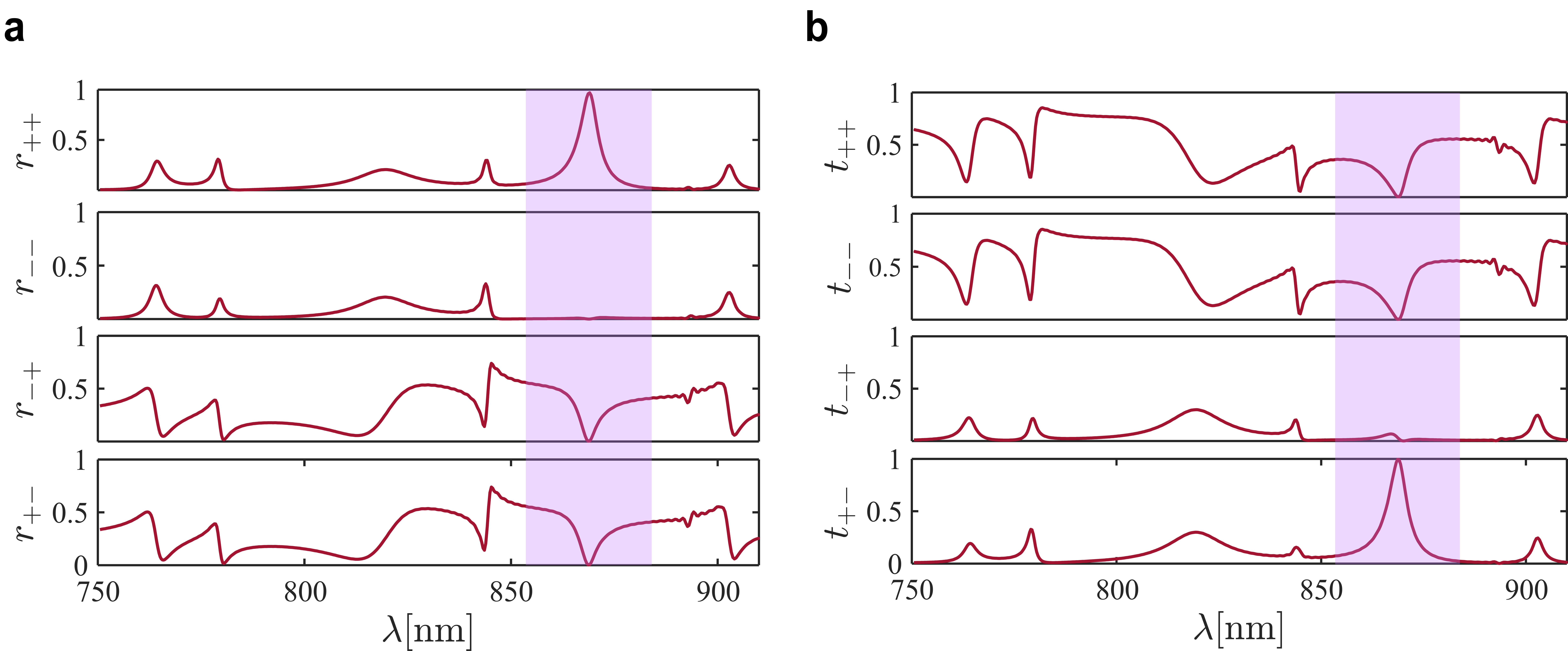}
\caption{\textbf{Simulation results for power reflection (a) and transmission (b) coefficients.}% \textbf{(a)} reflection \textbf{(b)} transmission. 
The subscripts $+$ and $-$ denote  RHCP and LHCP modes respectively.  The bandwidth of maximum chirality is shaded.}
\label{fig:Jones_matrix}
\end{figure}

The simulated polarization-resolved reflection and transmission coefficients are displayed in Fig.~\ref{fig:Jones_matrix}. The power reflection and transmission coefficients, denoted by $r_{ij}$ and $t_{ij}$, respectively, are related to the elements of Jones matrices as  $r_{ij}=\abs{\mathcal{R}_{ij}}^2$ and  $t_{ij}=\abs{\mathcal{T}_{ij}}^2$. The results confirm that over the operational frequency band (the shaded region in Fig.~\ref{fig:Jones_matrix}), the photonic crystal is maximally chiral \cite{2016_PRX_Fernandez}: at the target wavelength, it selectively reflects light and retain its handedness and, in compliance with symmetry constraints, transmits the opposite spin while flipping its helicity.

{This realization of maximum intrinsic chirality in  monolithic structure is exceedingly} surprising.  As pointed out earlier, the origin of the {chiroptical} effects can be traced back to the simultaneous excitation of effective in-plane magnetic and electric moments induced within the building blocks of the structure\cite{2017_NatLS_Zhu,2017_SR_chiralitytheory}.  Using the dipolar approximation, the optical response of the structure can be described on the basis of the net electric dipole per unit cell i.e. $\vp{p}=\frac{1}{i\omega}\iiint\vp{J}\mathrm{d}v$ and the net magnetic dipole moment calculated as $\vp{m}=\frac{1}{2c}\iiint \vp{r}\times\vp{J}\mathrm{d}v$, where $\vp{J}$, $\omega$ and $c$ are the polarization current, frequency, and the speed of light in vacuum, respectively. Since both the electric and magnetic dipole modes radiate primarily along the directions normal to their axis, their cooperative action requires a co-planar and co-linear excitation of the moments so that $\vp{p}_{||}.\vp{m}_{||}\neq 0$, where $\vp{p}_{||}$ and $\vp{m}_{||}$ refer to the components of the dipoles tangential to the plane of the slab \cite{2017_SR_chiralitytheory}. For this to occur, the slab should be thick enough so the polarization/displacement current can be circulated within the vertical cross section of the structure. In contrary to the common wisdom however, a giant circular dichroism with maximum chirality takes place in the PC slab whose thickness is much less than the target wavelength for operation.    

%%%%%%%%%%%%%%%%%%%%%%%%%%%%%%%%%%%%%%FIGURE
\begin{figure}[!t]
\centering
\includegraphics[width=1\textwidth]{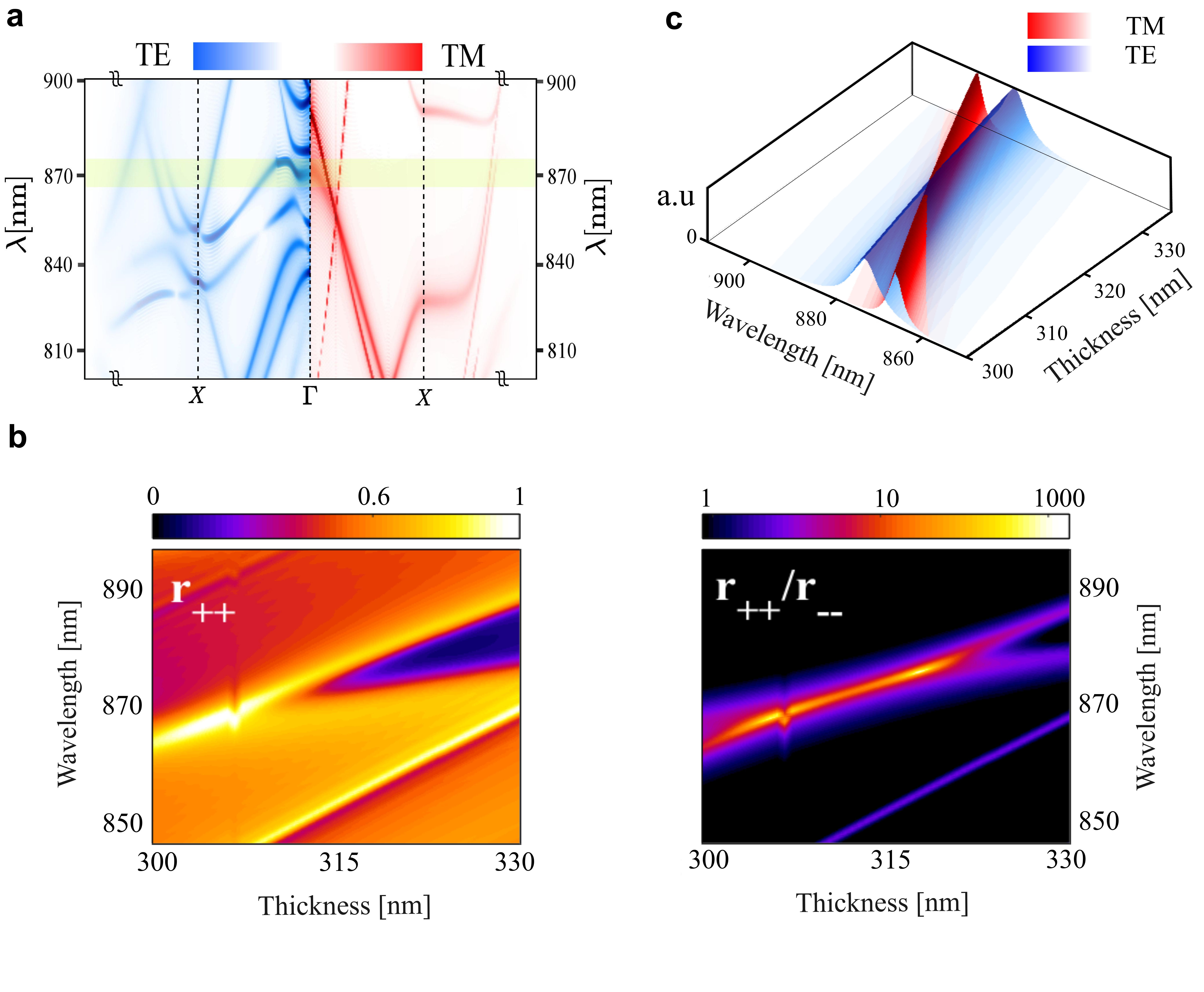}
\caption{\textbf{Band diagram and mode hybridization plots}. Two low-Q  TE and TM modes get hybridized by varying the thickness of the slab. For the thickness of $t\sim 309\mathrm{nm}$ the modes are properly hybridized guaranteeing the co-excitation of in-plane magnetic and electric dipole moments at the target wavelength of $\sim 870\mathrm{nm}$.  
\textbf{(a)} Relevant part of the band diagram for TE (blue) and TM (red) modes. The color axis displays the spectral function obtained from time-domain simulations. High-Q Bloch modes are brighter as they exhibit stronger resonance. The displayed part of the band is within the radiation continuum (above the light cone) and the yellow shaded region is the operation band of the mirror. 
\textbf{(b)}
Total reflectively of the chosen helicity, $r_{++}$ shown on the left and  extinction ratio $r_{++}/r_{--}$ shown on the right. The hybridization causes extreme chirality manifested in a large extinction ratio.
\textbf{(c)} Spectral distribution of the TE and TM modes. The distributions are normalized and they thus display the center wavelength associated with each Bloch mode as well as their line-width. The modes are extremely leaky  and thus having low radiation quality factors which in turn leads to GMR.}
\label{fig:Hybridization}
\end{figure}

%%%%%%%%%%%%%%%%%%%%%%%%%%%%%%%%%%%%%%%%%%%%

The key to understanding the operational principles of our chiral PC mirror is the Guided Mode Resonance (GMR). Here, the maximum intrinsic chirality is achieved by engineering low-Q TE-like and TM-like modes within the radiation continuum. Since the Bloch modes across the band-edges of the photonic crystal structure have low lateral expansion velocity, they just radiate back into free space and thus effectively act similarly to the Mie and  Fabry-P\'{e}rot resonant modes in {dielectric} metasurfaces and supercavities \cite{2017_PRL_Rybin}. 
The details of the geometry of the unit cell allow us to adjust the radiation quality factor associated with each virtual resonant mode so that they become extremely leaky. 
The leaky modes provide an efficient way to channel light from within the slab to the external radiation and, at the same time, the judicious design of the structure leads to formation of  TM modes that extend well outside of the slab. Thanks to the latter, desired magnetic moments can be excited. The TE-like modes can effectively generate the desired in-plane electric dipole moment and the in-plane magnetic dipole is generated by the TM-like modes within the radiation continuum. {If the slab dimensions are properly selected, the TE-like and the TM-like modes would acquire degenerate resonance frequencies which allows their hybridization. This in-turn  gives rise to the strong intrinsic chirality.} 
{Since the radiation channels associated with the TE-like and TM-like modes are not fully orthogonal, a \textit{via-continuum coupling} occurs, leading to a slight removal of degeneracy}\cite{2017_PRL_Rybin,2016_Nat_Hsu}.The relevant part of the band diagram around the band-edge i.e. $\vp{k}=0$ (denoted by $\Gamma$) is displayed in Fig.~\ref{fig:Hybridization}\textbf{a}. {
The band diagram is obtained by FDTD simulation with Bloch boundary condition on a single  unit-cell. This is simulated by placing random dipole sources in the unit-cell --so that all possible modes are excited-- and recording the field in time domain.  High-Q modes are less leaky and therefore upon their excitation, they last longer in time. The color axis displays spectral function obtained from the FDTD simulation; the modes having less radiation leakage exhibit stronger resonance and they are thus brighter.} Within the yellow shaded frequency band the desired {low-Q} TE and TM modes meet each other at the band-edge, which enables their hybridization. 
%and thus the PC mirror exhibits chirality.  

{The anomalous reflection of photonic crystals occurs due to the interference of the leaky Bloch modes and the continuum of unbounded modes and it thus exhibits  Fano-shape spectral profile} \cite{2003_JOASA_FAN}.{ The electromagnetic dipole moments in helical basis (which are commonly called $\sigma$-dipoles }\cite{2018_Optica_Eismann}){ consist of parallel magnetic and electric dipole moments of equal amplitudes that are phase shifted by $\pm \pi/2$. The presence of resonance based on $\sigma$-dipoles gives rise to the chiroptical effects of interest. At the design wavelength, the reflection from our PC mirror is generated by a resonant-coupling of circularly polarized light  to $\sigma$-dipoles which are essentially produced by the co-excitation of TE-like and TM-like modes as well as through the background reflection.}

To reveal how mode hybridization leads to chirality, we performed parameter tuning to decouple the modes. The thickness of the slab has been varied to observe the variations of the reflectively of the chosen helicity as well as the extinction ratio as the figures of merit. Fig.~\ref{fig:Hybridization} \textbf{b} 
show the investigated crossing region with reflectivity results from finely sampled simulations. The {spectral distribution} of the associated TE and TM modes, calculated through FEM method (ANSYS HFSS Inc.), are shown in Fig.~\ref{fig:Hybridization} \textbf{c}. {The distributions are normalized and they display the resonance wavelength and the line-width  associated with each Bloch mode.} To confirm that mode crossing takes place, we apply a mode tracing scheme. As expected, the TE-TM-mode crossing occurs around the predicted thickness.

%%%%%%%%%%%%%%%%%%%%%%%%%%%%%%%%%%%%%%%%%%%%%%%%%%
\begin{figure}[t]
\centering
\includegraphics[width=0.85\textwidth]{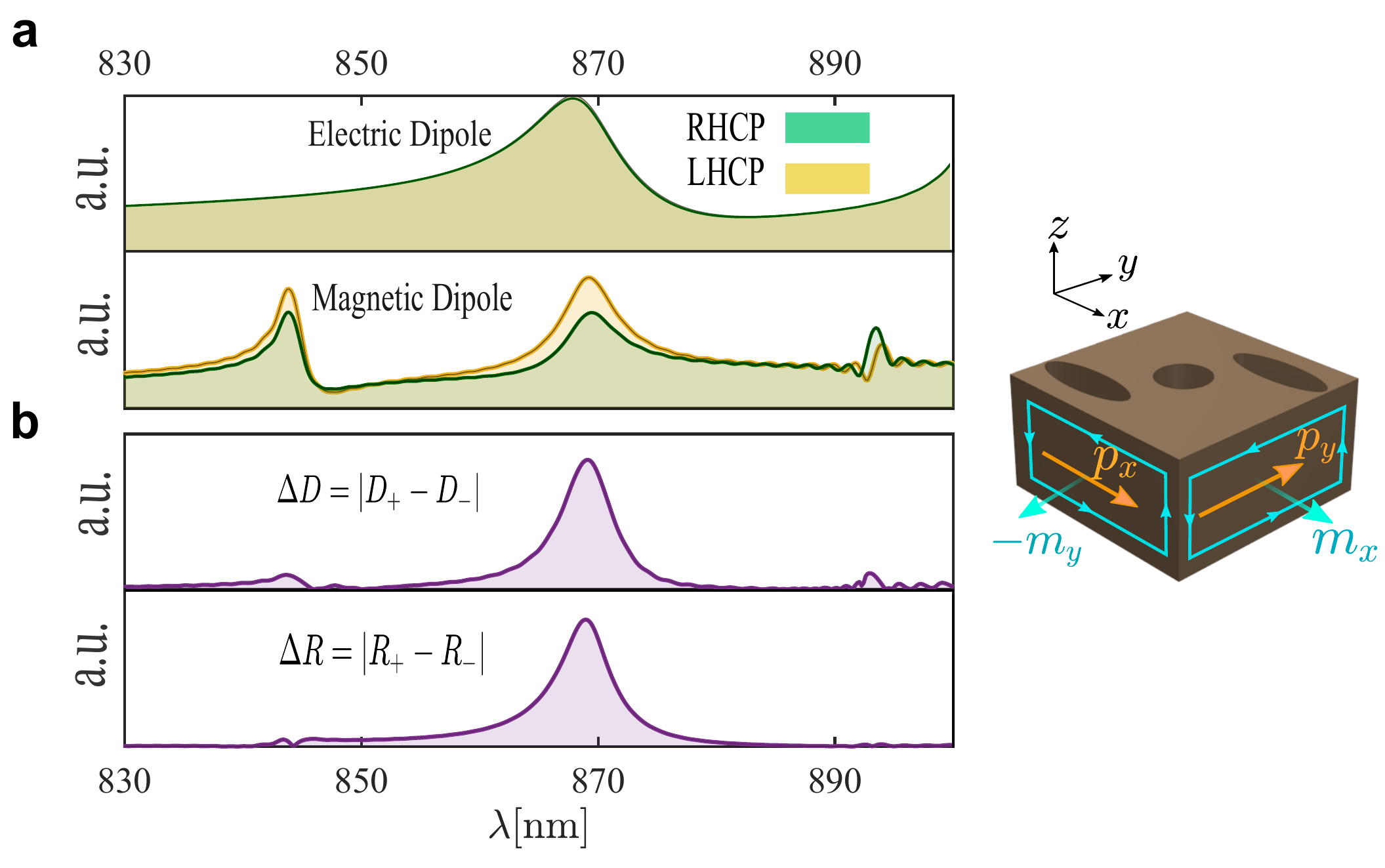}
\caption{\textbf{Spectral variations of the induced dipole moments.} \textbf{(a)} Dipole moments for RHCP and LHCP incidence (the green and yellow shaded curves respectively).   The in-plane electric dipole $\abs{\vp{p}_{\pm}}$ and the magnetic  $\abs{\vp{m}_{\pm}}$ are displayed in the upper and lower plots respectively. Here, the `+' and `-' subscripts denote RHCP and LHCP respectively. \textbf{(b)} Inner product of the in-plane electric and magnetic dipoles, i.e., {$D_{\pm}=\abs{\vp{p}_{\pm}\cdot \vp{m}_{\pm}}$} and the reflectivity contrast $\Delta R=\abs{R_+-R_-}$. The reflectivity contrast $\Delta R$ follows  $\Delta D=\abs{D_+-D_-}$.} 
\label{fig:dipoles}
\end{figure}
%%%%%%%%%%%%%%%%%%%%%%%%%%%%%%%%%%%%%%%%%%%%%%%%%%

To look more closely into the dipolar interpretation of our structure's predicted behavior, we studied the impact of  dipole interactions. We calculated the in-plane components of the induced electric dipole $\vp{p}$ and the magnetic dipole $\vp{m}$ per unit cell for incident light of both circular polarizations. The toroidal dipoles, as well as the higher order multipoles, were found to have negligible contributions here.
 Fig.~\ref{fig:dipoles}\textbf{a} displays the spectral distribution of the induced dipole moments. {The induced electric dipoles are identically excited for opposite helicities whereas the magnetic dipoles' strength are different over the chiral bands.}   Around the design wavelength, where modes are properly hybridized, the inner products of the resultant dipoles are distinctly different for the opposite helicities. Intriguingly, the circular dichroism at reflection follows magneto-electrical dipole interactions (see Fig.~\ref{fig:dipoles}\textbf{b}). 
 
{The origin of the difference in the induced magnetic dipoles shown in Fig.}~\ref{fig:dipoles}\textbf{a} {can be explained based on symmetry considerations. The induced polarization currents for circularly polarized incident fields can be naturally decomposed into a chiral and an achiral component namely $\vp{J}_c$ and $\vp{J}_a$ respectively. The chiral component $\vp{J}_c$ is induced differently for the opposite helicities and it  carries in-plane chiral symmetry in its distribution. Also, owing to the two-fold rotational symmetry, $\vp{J}_c$  has a definite parity under in-plane space inversion.  
Specifically, it is observed that the chiral polarization vector is an odd-parity vector field i.e. $\mathcal{P}_{xy}\{\vp{J}_c\}=-\vp{J}_c$ where $\mathcal{P}_{xy}$ is the in-plane parity operator. 
Since the electric dipole moment is obtained by the direct integration of the polarization current over the unit-cell, the net electric dipole moment loses its in-plane chiral features. Therefore, the electric dipole moments are equally excited for both helicities of the incident light.  In contrast, 
as the net magnetic dipole moment is calculated by the direct integration of $\vp{r}\times\vp{J}$ and 
$\mathcal{P}_{xy}\left\{\vp{r}\times \vp{J}_c\right\}=+\vp{r}\times \vp{J}_c$, the chiral polarization components should have non-vanishing contribution in the magnetic dipole moment which in turn leads to the excitation contrast observed in Fig.}~\ref{fig:dipoles}\textbf{a}.

{The magneto-electric dipole excitation described above is fundamentally different from  the operational mechanism of  loss-less planar dielectric metasurfaces made of high-refractive-index nanopillars. Analogous to  high contrast gratings that essentially operate in a dual-mode regime} \cite{2012_OpticsExpress_Karagodsky}{,  such all-dielectric metasurfaces support multiple guided-modes and formation of \textit{supermodes} in a symmetry-broken geometry can result in asymmetric transmission for orthogonal polarizations. Therefore, the nanopillars need to be sufficiently tall to accommodate internal multi-mode propagation. It has been recently demonstrated  that judicious design of nanopillars carrying in-plane geometrical chirality can potentially result in different coupling of opposite helicities to the waveguide-array modes which in turn leads to their differential response between left- and right-handed circularly polarized lights}\cite{Ma2018}. \citet{2017_PhysRevApplied_Ye} {have shown that multi-mode interference can also arise in metasurfaces made of low-loss metallic nanostructures with finite thickness. Consequently,  properly designed metallic nanoposts can support surface plasmon modes whose different interference schemes for opposite helicities  yields a giant chiroptical effect}\cite{2017_PhysRevApplied_Ye}.

\subsection{Experimental results}

The silicon nitride layer for our structures was grown on a silicon wafer using low-pressure chemical vapor deposition (LPCVD) producing a film with refractive index of 2.26 at the designed wavelength. The structures were fabricated through soft-mask electron beam lithography followed by plasma etching and a KOH undercut. {This process leaves us with a truly free-standing  photonic crystal membrane and the undercut area is sufficiently deep so that the impact of silicon substrate can be safely disregarded.}  A scanning electron microscope (SEM) image of the fabricated device is displayed in Fig.~\ref{fig:schematics}\textbf{c}. 

The experimental characterization of the photonic crystal mirror was carried out by free-space illumination with {a beam of supercontinuum white-light laser}. The beam is focused onto the PC sample through a low numerical aperture objective lens to assure the wave-front of the excitation remains similar to a plane wave. { To obtain the reflection spectrum, we collected the reflected light into a spectrometer via a single-mode fiber.} {The focused beam at the sample was approximated by a Gaussian profile;  the corresponding beam waist at the sample was estimated to be $w_0\approx 18\mathrm{\mu m}$}. To get rid of artifacts originating from unwanted rays, the reflected beam passes through a confocal reflectometry setup with appropriate polarimetric arrangements. The setup is designed to monitor the four components of the power reflection matrix in circular basis: $r_{++}$, $r_{--}$, $r_{+-}$, and $r_{-+}$. The confocal configuration is necessary to compensate for the long depth-of-field associated with the loosely focused beam so the rays reflecting off the undercut area are largely avoided. To further reduce interference of the rays reflecting  from the thick silicon layer underneath the PC membrane, 
the undercut region has a V-shaped cross section in the silicon substrate and thus reflections from its walls are mainly off-normal. Additional details of the optical setup can be found in the Supplemental Material.

%%%%%%%%%%%%%%%%%%%%%%%%%%%%%%%%%%%%%%%%
\begin{figure}[!t]
\centering
\includegraphics[width=1\textwidth]{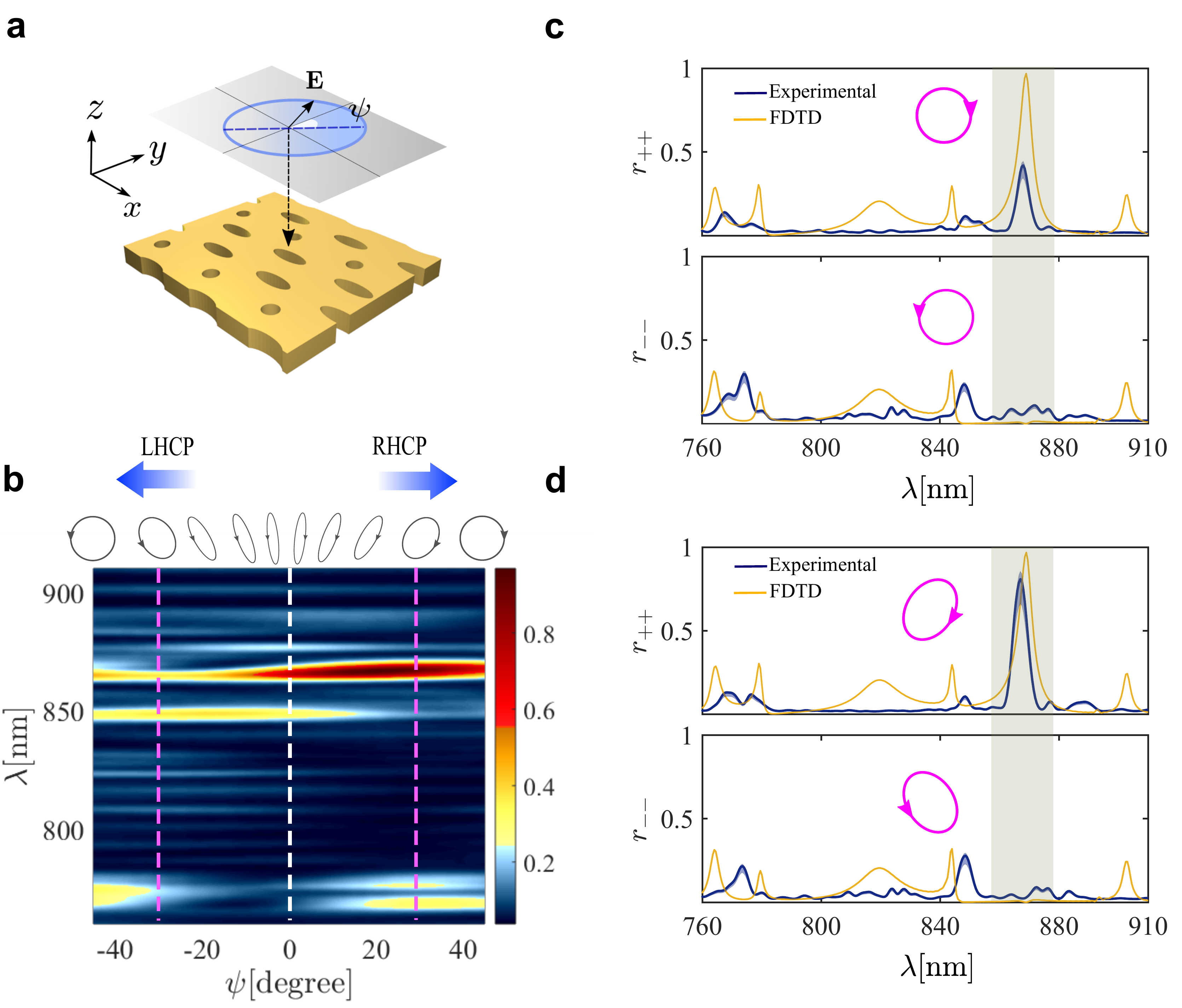}
\caption{\textbf{Experimental results:} \textbf{(a)} incident wave's polarization ellipse generated by the quarter wave plate placed before the focusing objective lens. The major axis is tilted  by the angle of $\psi$  with respect to the polarization of the input beam. The axial ratio is $\tan\psi$. For $\psi=\pm 45^{\circ}$ the incident light is circularly polarized. \textbf{(b)} A color plot of the measured {normalized} reflection spectrum for different elliptically polarized beams. Around the design wavelength, the eigen-polarizations that exhibit maximum reflectivity and chirality are elliptical.   \textbf{(c)} Reflectivity of a circularly polarized incident light to the same helicity. Due to fabrication imperfections, the measured reflectively is relatively poor (only $\sim50 \%$). \textbf{(d)} reflection spectrum of elliptically polarised light with $\psi\sim \pm 30^\circ$ (for right handed and left-handed elliptically polarized light) to the same polarization. The extinction ratio  at the wavelength of $\sim 868\mathrm{nm}$ is $\frac{r_{++}}{r_{--}}\sim 33$. }
\label{fig:experimental}
\end{figure}
%%%%%%%%%%%%%%%%%%%%%%%%%%%%%%%%%%%%%%%%%%%%%%%%%%
 
Due to a slight astigmatism of the electron beam during the lithography process, as well as imperfect etching, the fabricated samples exhibit some anisotropy that entails a modest  performance degradation of the PC mirror. Specifically, it was observed that the eigen-polarizations are not purely circular. To experimentally explore this effect, we carried out a  polarization-dependent reflectometry. The polarization of the incident light is adjusted by means of a broad-band quarter-wave plate placed before the objective lens. This allows us to explore a wide range of elliptically polarized incident light. The total reflectivity of the PC mirror for different states of polarization is shown in Fig.~\ref{fig:experimental}\textbf{b}. Note that all reflectivity measurements are calibrated based on the known reflectivity of the un-patterned silicon nitride on a silicon substrate. It is observed that around the target wavelength of $\sim 870\mathrm{nm}$, the sample exhibits extreme chirality.  However, near-unity reflection occurs for a right-handed  elliptically polarized light with an axial ratio of $AR=\tan 32^\circ\approx 2:3$ (The dark red spot in ~\ref{fig:experimental}\textbf{b}). We emphasize that owing to the resonant nature of the chiral reflection mechanism, sensitivity to any structural deformation is expected to be pronounced. However, refinements to our fabrication procedure should allow us to make the eigen-modes purely circular. 
%As mentioned previously, the discrepancies between theory and experiment are most likely caused by fabrication imperfections. 
The observed deviation of the eigen-polarizations from circular to elliptical indicates a cross-coupling between the opposite spins. To inversely reconstruct the actually fabricated device, we carried out a diagnostic analysis based adjoin shape optimization \cite{2018_Nature_Molesky}. This revealed that due to the imperfect etching, walls of the photonic-crystal holes are not perfectly vertical so the diameters of the holes at the bottom and top surface are slightly different. This small imperfection was observed in a zoomed SEM image of the device as two concentric  boundaries appear around the individual holes (see Supplemental Material). In agreement with intuition, mirror symmetry-breaking in z-direction causes {an additional} cross-coupling between the associated TE-like and TM-like modes which in turn leads to the presence  of the off-diagonal elements $\mathcal{R}_{+-}$ and  $\mathcal{R}_{-+}$ in the reflection matrix.

 The measured spin-preserving components of the reflectivity tensor for circular and elliptical polarziations are shown in Fig.~\ref{fig:experimental} \textbf{c} and \textbf{d}, respectively. The off-diagonal elements are less-pronounced within the chiral band and they are presented in the Supplemental Material.  There are some discrepancies between simulations and experiment including a few nanometer wavelength shift of the dominant features of the simulated and experimentally observed spectra. However, when overlaying the simulations and experimental results, we observe a very good agreement overall and the slight differences most likely arise from fabrication imperfections. The experimental results confirm selective reflection of the incident light for two elliptically-polarized eigen-modes with opposite helicity with an extinction ratio that can reach up to $r_{++}/r_{--}\gsim 30$. Such extreme chirality is unprecedented among previously reported experiments. The blue shaded reliability curves account for uncertainties in calibration of the reflectivity measurement results including a slight loss of coupling to the single mode fiber when the illuminated spot is moved to the un-patterned area. {It is also worth pointing out that the slightly smaller measured reflectivity compared to the simulation results can be partly attributed to the finite size of the focused beam at the sample. As is further evidenced by a plane-wave expansion analysis presented in Supplemental Material, the Gaussian beam used in our experiment contains obliquely incident plane-waves which have significantly lower reflection coefficients and thus even in an ideal scenario, the reflectivity of such Gaussian beam cannot exceed $\sim 85\%$.}

\begin{figure}[t]
\centering
\includegraphics[width=1\textwidth]{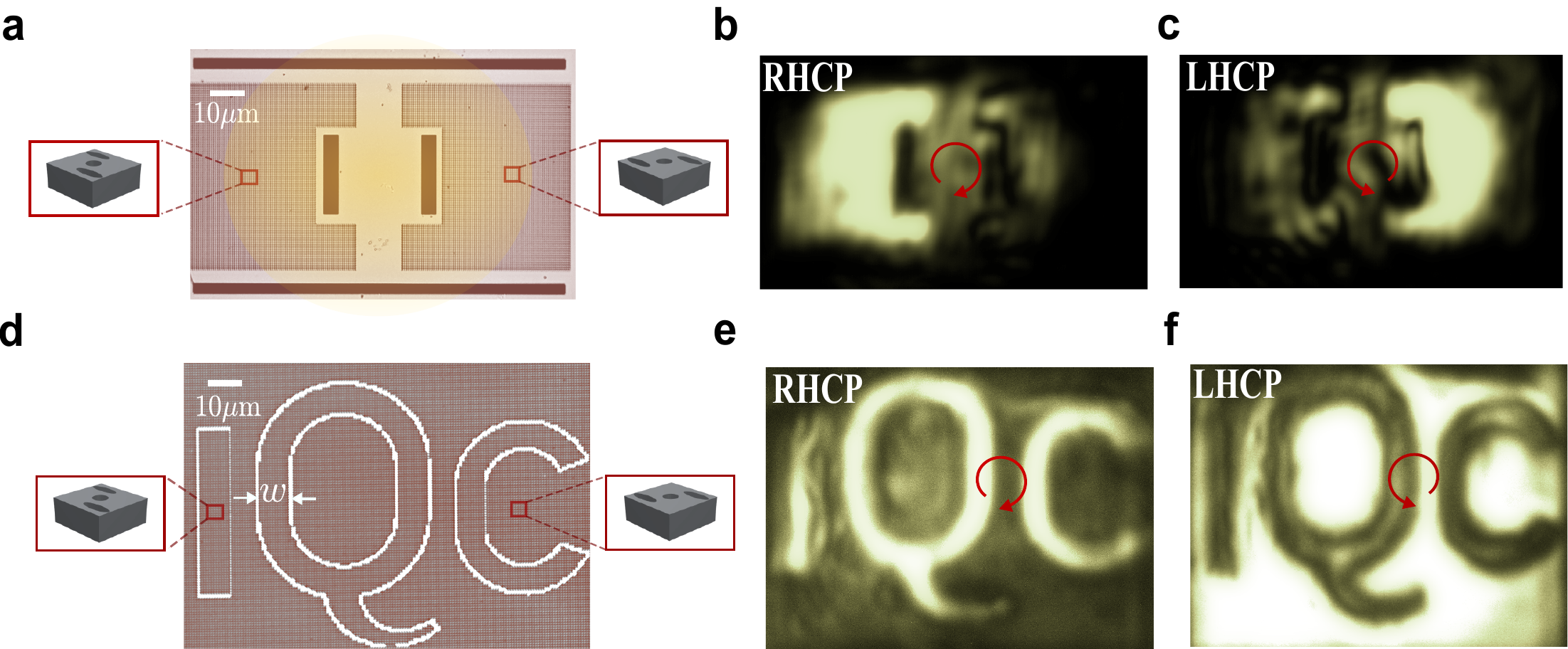}
\caption{\textbf{Reflection of the apposite enantiomers upon circularly polarized illuminations.} \textbf{(a,b,c)} Imaging using a focused and spatially coherent laser source. \textbf{(a)} Optical microscope image of the fabricated pattern. The left `C' structure reflects RHCP while the `mirror C' structure on the right  only reflects LHCP. Each C is $\sim 50\mathrm{\mu m} \times 50\mathrm{\mu m}$. The yellow shaded circle shows the region illuminated by the coherent laser beam \textbf{(b,c)} reflection of CP light to its co-circular polarization  \textbf{(d,e,f)} Imaging using a laser beam with a scrambled wavefront.
The interior and exterior of the letters (IQC) are made of photonic crystals with opposite chiral patterns.\textbf{(d)} Optical microscope image of the fabricated IQC pattern. The letters are about $10\mathrm{\mu m}$ wide i.e. $w\sim 10\mathrm{\mu m}$ . \textbf{(e)} $\mathrm{RHCP}\to\mathrm{RHCP}$  and  \textbf{(f)} $\mathrm{LHCP}\to\mathrm{LHCP}$ reflection imaging. In both cases, extreme chirality occurs at the wavelength of $\sim 824\mathrm{nm}$ so is the wavelength of the laser source.  }
\label{fig:imageing}
\end{figure}

To demonstrate the robust  performance of the chiral mirrors, we performed two experiments involving polarization-resolved imaging. In the first experiment, two C-shaped photonic crystal membranes with opposite chiral patterns (two enantiomeric configurations) were fabricated. We identified the chiral operational band of the PC sample  through polarization resolved spectroscopy. An optical microscope image of the fabricated sample is displayed in Fig.~\ref{fig:imageing}\textbf{a}. 
%Having performed polarization resolved  spectroscopy, the chiral operational band of the PC sample was identified. 
The pattern was then illuminated by a monochromatic and spatially coherent laser beam from a tunable continuous-wave Ti:Sapph laser at $824\mathrm{nm}$ -- the wavelength at which the PC sample exhibited extreme chirality. %Since, in this experiment, the sample is exposed to a spatially coherent beam, the illumination is fairly a plane wave. 
The polarization resolved images are shown in Figs.~\ref{fig:imageing}\textbf{b} and \textbf{c}. The images have been captured after a circular-polarization filter collecting only spin-preserving (co-circularly polarized) reflection and thus artifacts originating from the background are largely suppressed. Under purely circularly polarized illumination, only one of the photonic crystal structures looks bright, with the fringing in the image arising from the high level of spatial coherence of the illumination in this case. 
%As mentioned earlier, in compliance  with the intuition, regular background reflections or Snell reflection from the dielctric interfaces would involve flipping the spin and thus they do not appear in the the final image.  

We also created a pattern of letters (IQC) shown in Fig.~\ref{fig:imageing}\textbf{d}. The interior and exterior of the letters are made of photonic crystal structures with opposite enantiomeric patterns. Following the experimental procedure outlined above, we again performed a polarization-resolved imaging. This time, however, we illuminated the pattern with a monochromatic but spatially incoherent laser beam. To suppress the beam's spatial but not temporal coherence, the laser beam is first focused onto a rotating ground glass diffusers and collimated again \cite{2017_OSA_Stangner}. As can be seen in Figs.~\ref{fig:imageing} \textbf{e} \& \textbf{f}, the letters look bright with a clear contrast with respect to the background under RHCP illumination, while for LHCP illumination, the exterior region looks bright and the letters are dark. Intriguingly, the contrast between complementary domains is relatively high, even around the sharp edges. Since the letters are about $10\mathrm{\mu m}$ wide, this demonstrates that strong chirality can be achieved even for photonic crystal structures which are periodic over small scale areas. {This observation is consistent with the results obtained from full-wave simulations of a number of finite-size photonic crystal slabs presented in Supplemental Material.}
%the periodicity of the photonic crystal structure over letters is refuted; the chirality is still strong though.        

\section{{Discussion}}

Compared to the state-of-the-art chiral metasurfaces, including the ones made of complicated 3D chiral shapes \cite{2009_IOP_Optics_Plum,2019-ACS-Rajaei,2017_NanoLett_Chen}, our PC mirror exhibits a remarkably large optical chirality despite its structural simplicity. The reflectivity of the chosen helicity to the same state of polarization can, in the ideal limit, reach up to $\sim100\%$ and we observed about  $\sim 80\%$ reflectivity with an extinction ratio going beyond $30:1$ in our fabricated devices.
In contrast, the majority of the experimental demonstrations of {chiroptical} effects in visible and near infrared range have reported circular dichroism in
transmission\cite{2017_NatLS_Zhu,2017_NanoLett_Chen,2017_SR_Hu,2019-NanoLett-Yang,2017_science_Hentschel, 2012-NatCom-Zhao,2012_PRB_Zhou,2014_NaoLett_Cui,2018_NatLS_CHEN_SPIN,2018_ACS_Wu,2005_PRL_KuwataGonokami}. Among these, the most notable ones include 
a demonstration of  a chiral metasurface by \citet{2017_NatLS_Zhu}, which achieved spin filtering with an extinction ratio of $\sim 9:1$ while almost $\sim 90\%$ of light with the selected helicity was transmitted, and  a planarized chiral structure made of multiple layers of twisted metamaterials by \citet{2012-NatCom-Zhao}. The latter structure exhibits almost $\sim 30\%$ transmission difference (and the extinction ratio of $\sim3:1 $) for the opposite helicities. Most recently, a double-layer plasmonic chiral structure was proposed by \citet{2018_NatLS_CHEN_SPIN}. Although the structure enables filtering the opposite spins with a large enough extinction ratio, the maximum transmission of the chosen helicity at the chiral band remains below $\sim8\%$.  The most notable experimental demonstrations of nano-scale chiral spin-preserving mirror are presented by \citet{2017_NanoLett_Kang} {and} \citet{2017_PhysRevApplied_Ye}. The reported experimental results\cite{2017_NanoLett_Kang} indicate the reflectivities of  $\sim 80\%$ and $\sim 20\%$ for the intended and rejected helicities respectively. Our PC mirror outperforms most of the chiral mirrors realized to date in most categories, although its operational mechanism introduces some sensitivity to the angle of incidence and its resonant nature gives rise to a relatively narrow operation band and a certain susceptibility to fabrication imperfections. However, the structure can be further optimized for broadband operation \cite{2017_ACS_Flannery} which remains as the next-step study for further developments.{ At the same time, there is a number of applications such as gas lasers for which a narrow operation band is not really a liability and can even be viewed as an asset for improved coherent operations.} 
It is worth pointing out that the maximum angle of view is limited by the region of the k-space where the group velocities of the associated Bloch modes remain negligibly small. According to Fig.~\ref{fig:Hybridization}\textbf{a}, the dispersion of the TE and TM Bloch modes (within the shaded area) is fairly flat  over only a limited region within the k-space and so is the angle of view. Band flattening techniques \cite{2004_IOP_Cond_Takeda} may  thus be employed to design a chiral PC mirror that is less sensitive to the angle of incidence.

In summary, we have designed, fabricated and experimentally demonstrated an intrinsically chiral photonic crystal mirror that upon normal illumination selectively reflects circularly polarized light without reversing its handedness. Guided mode resonance due to the interplay of TE and TM modes across the band-edge leads to such  enormously strong chirality with near unity reflectivity contrast. Although the structure exhibits certain sensitivity to the angle of incidence, %even for non-plane wave incidence, 
its extreme chirality renders it a compelling device compared to the seminal works in this area of nanophotonics. 
%This was examined through two experiments involving nano-scale imaging with the objects made of our PC mirror.   

\section{Methods}
\subsection{Sample fabrication} 
The photonic crystal membranes are fabricated from silicon nitride, a material that provides low absorption in the near-IR region for which the mirrors are designed to be highly reflective.  The silicon nitride is grown on a 4 inch silicon wafer by low-pressure chemical vapor deposition (LPCVD) to produce a film with a refractive index of 2.26 at 850nm. The wafer is diced into $8\times8$ mm chips using a thick ($\sim 1\mu$m) protective layer of PMMA.

Since the exact thickness of the silicon nitride is not always consistent and can have spatial variance within the wafer, we grow the silicon nitride to be larger than the desired membrane thickness. We then remove the protective PMMA layer with Remover PG at an 80$^\circ$C bath with sonication and use reactive ion etching (RIE) to etch the silicon nitride on a given chip to the correct thickness. The plasma etching recipe used is a 130/80 sccm mixture of C$_4$F$_8$/SF$_6$ at 10mTorr, with a 1000 W ICP RF power, 30 W platen RF power, and platen temperature set to $15^\circ$C. Before running any etching process, we first condition the chamber with this etch recipe for 45 minutes. The silicon nitride etching rates before each run are found by running the process for 1 minute on a test chip and then measuring the silicon nitride thickness before and after the etching using filmetrics. Typical etch rates are 15-20 nm/min.

The fabrication of the PC pattern is done by electron-beam lithography in combination with dry and wet etching processes. We begin by spinning ZEP520A (Zeon Chemicals) positive resist to a thickness of $\sim$700 nm using a spin speed of 1500 rpm for 60 seconds at a ramp rate of 3000 rpm/min and baked for 2 minutes at $180^\circ$C. The resist is developed using amyl acetate for 90 seconds after being exposed with the PC pattern by e-beam lithography at 100 keV. Transfer of the pattern from the e-beam resist to the silicon nitride is achieved again by RIE plasma etching using the previously described silicon nitride recipe. The sample is etched to $125\%$ of the thickness to ensure complete penetration through the film leaving perpendicular side walls.

In order to reduce back-reflections from the silicon surface, we perform a silicon undercut using 45$\%$ KOH solution at 80$^{\circ}$C, allowing for the patterned regions of the silicon nitride to become free-standing films. Typically the silicon wet etch is performed for $\sim 1$ hour followed by immersing the sample in two beakers of deionized (DI) water for 5 minutes each to neutralized the KOH.  These neutralizing DI baths are also held at 80$^{\circ}$C in order to prevent the formation of crystals on the surface of the sample. The samples are then transferred into two solutions of IPA for at least 5 minutes each to remove the DI water in the features of the pattern. This provides a surrounding solvent with a much lower surface tension, reducing the risk of breaking or cracking the delicate free-standing PC patterned films when finally dried with N$_2$ gas.

\subsection{Measurement procedure} 

The microscope setup (shown in Supplemental Material, Fig.~S4) employs a 5$\times$ objective  with a numerical aperture of $NA=0.1$ and a lens system which is consist of  two confocal lenses  (Thorlabs AC254-200-B and AC254-150-B) with focal distances of 20 cm  and 15 cm, respectively. The diameter of the focused beam  at the PC sample was measured by a sharp blade mounted on a high resolution translation stage. We observed a beam waist of $w_0\approx 18 \mathrm{\mu m}$ with no noticeable chromatic aberration between $600 \mathrm{nm}$ to $920\mathrm{nm}$ wavelength range. To make sure that the incident beam is precisely  normal to the PC mirror, we initially align the beam without the objective so the reflected beam is co-linear with the incident one.
The incoming beam is vertically polarized  and passes through a non-polarizing beam splitter (Thorlabs BS014).  To extract the four components of the Jones matrix in the circular basis, we used a zero-order achromatic $\mathrm{\lambda/4}$-plate (Thorlabs AQWP05M-980) right before  the objective and after the  non-polarizing beam splitter. This ensures that wave-front is not distorted and eliminates the need for further calibration of the wave-plate. For $\theta=\pm45^\circ$ (where $\theta$ is the angle of the fast axis of the quarter wave-plate with respect to the vertical direction), the PC mirror is illuminated by RHCP and LHCP light. After a round trip of propagation , the diagonal elements of the Jones matrix are coupled to vertical polarization and off-diagonal elements are coupled into the horizontal polarization at the output. We used  a zero-order achromatic $\mathrm{\lambda/2}$-plate (Thorlabs AHWP05M-980) together with a polarizer to selectively couple the reflected beam into a single mode fiber which is connected to the spectrometer.

\section{Acknowledgements}
This research was undertaken thanks in part to funding from the Canada First Research Excellence Fund. Additionally, this work was supported by Industry Canada, NSERC Discovery grant, and by Ontario's Ministry of Innovation Early Researcher Award. The authors also thank the Quantum Nanofab facility for the support with the nanofabrication part of this work and, in particular, Nathan Nelson-Fitzpatrick for growing the LPCVD SiN films.

\section*{Author contributions}
B.S., J.F. and M.B. defined the project and conceived the experiment(s). B.S. conducted the experiments.  J.F. fabricated the device. B.S. developed the theoretical model and carried out the simulations.   Under the guidance of M.B.,  B.S and J.F. and R.A. wrote the manuscript and all authors reviewed the manuscript.
\section*{Additional information}

\textbf{Competing interests}:
The authors declare no competing interests
\newline
\textbf{Data availability:} Experimental data presented in this article is available from the corresponding author upon request.

\bibliography{main}

\end{document}